*sensors*

Article

# High Sensitivity real-time VOCs monitoring in air through FTIR Spectroscopy using a Multipass Gas Cell Setup

Annalisa D'Arco[1,2], Tiziana Mancini[2,3], Maria Chiara Paolozzi[4], Salvatore Macis[2,3], Augusto Marcelli[2,5], Massimo Petrarca[3,6], Francesco Radica[7], Giovanna Tranfo[8], Stefano Lupi[2,3] and Giancarlo Della Ventura[4,9,*]

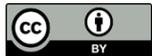

1   INFN-LNF Laboratori Nazionali Frascati, Via E. Fermi 54, 00044 Frascati, Italy
    augusto.marcelli@lnf.infn.it, annalisa.darco@roma1.infn.it
2   Department of Physics, University of Rome 'La Sapienza', P.le A. Moro 2, 00185, Rome, Italy
    tiziana.mancini@uniroma1.it, stefano.lupi@uniroma1.it, salvatore.macis@uniroma1.it
3   National Institute for Nuclear Physics section Rome1, P.le A. Moro 2, 00185, Rome, Italy
    stefano.lupi@roma1.infn.it
4   Department of Science, University Rome Tre, V.le G. Marconi 446, 00146, Rome, Italy
    giancarlo.dellaventura@uniroma3.it paolozzi.1698739@studenti.uniroma1.it
5   Rome International Centre for Materials Science Superstipes, Via dei Sabelli 119A, 00185, Rome, Italy
6   SBAI, Department of Basic and Applied Sciences for Engineering, University of Rome 'Sapienza', Via Scarpa 16, 00161 Rome, Italy  massimo.petrarca@uniroma1.it
7   Department of Engineering and Geology, University G. d'Annunzio Chieti-Pescara,
    francesco.radica@unich.it
8   Department of Occupational and Environmental Medicine, Epidemiology and Hygiene, INAIL, Monte Porzio Catone, 0078 Rome, Italy g.tranfo@inail.it
9   INGV, Via di Vigna Murata 605,00143, Rome, Italy

*   Correspondence: e-mail@e-mail.com; Tel.: (optional; include country code; if there are multiple corresponding authors, add author initials)

**Abstract:** Human exposure to Volatile Organic Compounds (VOCs) and their presence in indoor and working environments is recognized as a serious health risk, causing impairment of varying severity. Different detecting systems able to monitor VOCs are available in the market, however they have significant limitations for both sensitivity and chemical discrimination capability. During the last years we studied systematically the use of Fourier Transform Infrared Spectroscopy (FTIR) spectroscopy as an alternative, powerful tool for quantifying VOCs in air. We calibrated the method for a set of compounds (styrene, acetone, ethanol and isopropanol) by using both laboratory and portable infrared spectrometers. The aim was to develop a new, real time and highly sensitive sensor system for VOCs monitoring. In this paper, we improve the setup performance testing the feasibility of using a multipass cell with the aim of extending the sensitivity of this sensor system down to the part per milion (ppb) level. Considering that multipass cells are now available also for portable instruments, this study opens the road for the design of new high-resolution devices for environmental monitoring.

**Keywords:** VOCs; FTIR; sensor; accuracy; $ppm_v$; styrene; isopropanol; acetone; ethanol

## 1. Introduction

The term volatile organic compounds refers to a group of toxic organic chemicals which have low boiling points and evaporate easily at room temperature (RT), therefore they can rapidly spread in the environment, causing outdoor and indoor air pollution [1-4]. These compounds can be produced from natural and anthropogenic processes, such as human and animal metabolic processes, plant and tree emissions, forest fires, biomass and carbon combustion, but also from industrial and domestic activities including printing, building and storage of materials, food extraction and cooking [5-6]. Several studies show that in the worst cases exposure to VOCs, which can occur through inhalation, ingestion and dermal contact, may cause respiratory illness, neurocognitive impairment and cancer [7-8]. Due to their toxicity and the large spread of these chemicals in indoor and outdoor environments, the design of high-resolution, portable and reliable air quality monitoring tools is an important task to prevent VOCs overexposure, in particular in indoor environments such as working sites.

Nowadays, different analytical tools can be used for detecting VOCs in the atmosphere, i.e. gas semiconductor sensors [9-14], micromechanical resonant sensors [15] with high sensitivity, but also analytical instruments based on Photo-Ionization-Detectors (PIDs). Nevertheless, these devices do not allow easy discrimination among different chemical agents. At variance, analytical methods based on vibrational spectroscopies such as Terahertez (THz) and Infrared (IR) based techniques are able to accomplish this goal allowing both the detection of VOCs at very low concentrations and their reliable discrimination [10][16-18]. In addition, the ability of vibrational spectroscopy to analyze ambient air in real-time represents a unique feature, and this original approach is suitable for implementing portable devices. Notably, the potential of Mid-Infrared (MIR) spectroscopy is related to the numerous absorption molecular lines of interest allowing to recognize gas-phase VOCs. This ensures high selectivity and sensitivity for ambient-level detection (typically in the ppb range) of common hazardous air pollutants, which are typically strong IR absorbers.

In this research, we use FTIR spectroscopy in the MIR to selectively identify organic chemical compounds in air, starting from our previous study [18] where the general methodology was set up. After the first experiments by combining a dedicated PID sensor with MIR spectroscopy, we calibrated the IR technique for the quantitative analysis of a series of VOCs [18]: styrene, acetone, ethanol and isopropanol. Here, we extend the sensitivity of this technique by using a multipass gas cell. More specifically, the main aim of this work is to demonstrate the feasibility of gas monitoring down the ppb domain. The final goal is at develop a robust quantitative tool to be implemented on portable devices.

The availability on the market of compact multipass cells tailored to fit also on a portable IR spectrometer, similar to that described in [18] has opened the possibility for new monitoring strategies. To this purpose, we test here a laboratory-based gas cell to calibrate the spectroscopic analysis of four model compounds. Binary and ternary mixtures were also characterized to address the discriminating capability of the method when different VOCs are present in the atmosphere.

## 2. Materials and Methods

### 2.1. Materials

We considered four liquid VOCs: styrene, acetone, ethanol and isopropanol purchased by Sigma Aldrich and Carlo Erba. Styrene ($C_8H_8$ – Purity ≥99.0% Carlo Erba) belongs to the family of aromatic hydrocarbons, and can be found in varnishes, detergents, propellants and in the production process of plastic and packaging. Acetone ($C_3H_6O$ – Purity 99.5% Sigma Aldrich) is a ketone, commonly used as solvent in several industrial applications, such as polymer production, cosmetics, lacquers, cellulose acetate and varnishes. Finally, ethanol ($C_2H_6O$ – Purity ≥99.8% Sigma Aldrich) and isopropanol ($C_3H_8O$ – Purity ≥99.9% Sigma Aldrich) are alcohols and are both used as disinfectants and detergents, as additive solvents for cleaning optical and electronic components, and in the cosmetic field.

### 2.2. Method

FTIR measurements for the calibration of the multipass device were performed using the Bruker Vertex 70V interferometer at the Physics Department of Sapienza University in Rome. The schematic experimental setup is shown in Figure 1. It is similar to the portable device described in [18], with the exception of the gas cell, that in the present case has been replaced by the GEMINI gas cell with a nominal path length of 10 m and a volume of 2 l (http://www.gascell.com/). This multipass cell, allowing us to achieve a high sensitivity in the absorbance measurements, was connected to a sealed evaporation chamber, where a commercial Photo-Ionization Detector (PID) sensor was installed for real-time monitoring of the evaporated VOC. The PID sensor (TA-2100 Styrene Detector from Mil-Ram Technology, Inc., www.mil-ram.com) is calibrated for the detection of styrene in the range 1-100 ppm$_v$, where ppm$_v$ means parts per million volume. According to the manufacturer, this sensor has a sensitivity of 1 ppm$_v$. Concerning the other VOCs, the reading of ppm$_v$ was performed with PID sensor normalizing to a correction factor (CF from RAE, 2013 for a PID UV lamp at 10.6 eV) [19]. In this work, we used the following values for CF: 2.75 for acetone, 30 for ethanol, 15 for isopropanol and 1 for styrene. For these experiments, different amounts of liquid styrene, acetone, ethanol and isopropanol were introduced with a micropipette inside the evaporation chamber whose volume is ~ 0.6 l. The concentration was monitored against time using the PID sensor interfaced with a computer using a LabVIEW™ software. As soon as the countings provided by the PID were constant, indicating the attainment of the equilibrium condition within the chamber, corresponding to the maximum evaporation of the liquid, the left valve in Figure 1 was opened and the gas transferred into the multipass gas cell inside the spectrometer. The flow was ensured by pre-evacuating the cell (vacuum pressure around a few mbar) closed by two 25 mm wide and 4 mm thick KBr windows, using a vacuum system (Edwards T-Station 85) constituted by a turbomolecular and a diaphragm vacuum pump. Spectra collection was started immediately after opening the connection between the expansion chamber and the multipass, together with the simultaneous recording of the PID readings. A total of 15 spectra were collected during the gas distribution in the whole volume, each one being the average of 64 scans, with a nominal resolution of 2 cm$^{-1}$ in the 400-5000 cm$^{-1}$ frequency range. After few minutes spectra reach a constant intensity, i.e., indicating an equilibrium condition between the expansion chamber and the cell, therefore ten collected spectra were considered for the statistical data analysis Basic operations such as baseline, spectral windows identification and peak integration were performed using OPUS™ 8.2 software. The linear fit to determine the calibration curves was performed with the ORIGIN PRO™ software.

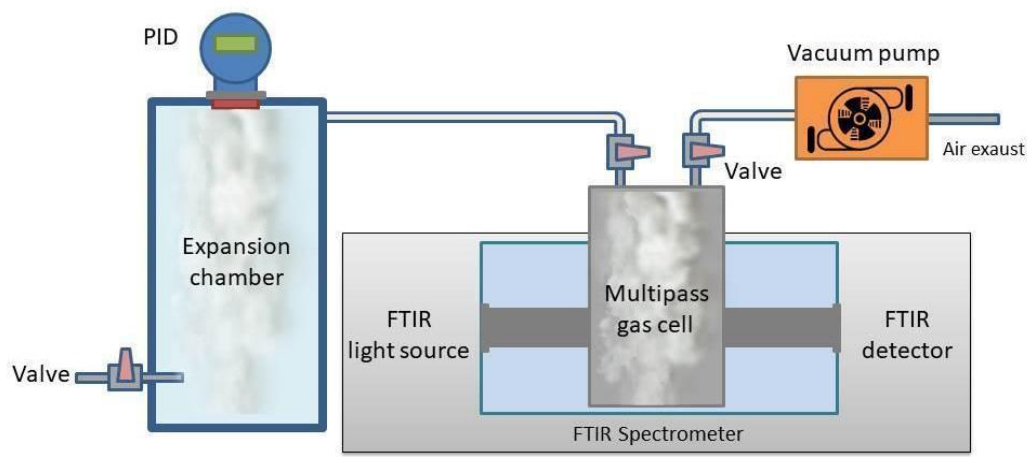

**Figure 1**. Schematic view of the multipass gas cell setup used for the calibration experiments. The multipass gas cell, located in the sample compartment of a Vertex 70v Michelson interferometer, was connected to a sealed evaporation chamber. The PID system was installed on top of the evaporation chamber and was connected to a computer for real-time monitoring of evaporated VOCs.

## 3. Results and Discussion

### 3.1. Characterization and statistic analysis

Figure 2 shows selected MIR absorption spectra of the investigated VOCs. The resulting spectra usually display several absorptions related to atmospheric $H_2O$ and $CO_2$ in the 3000-4000 $cm^{-1}$ and 1750-2100 $cm^{-1}$ ranges and around 2400 $cm^{-1}$, respectively. Beside these absorptions, in the spectral windows around 1100 and 3100 $cm^{-1}$ spectra show exclusively the typical absorption features characteristic of each VOC. In Figure 2, the region between 750 and 1300 $cm^{-1}$ is highlighted together with the features selected as representative for each compound during the calibration. For styrene the peak at 910 $cm^{-1}$ is assigned to an out of plane bending of CH bonds in the aromatic ring [20-21] while in the same spectral range acetone shows only a broad absorption centered at 1229 $cm^{-1}$ due to $CC_2$ antisymmetric stretching [22]. The ethanol band is centered between 1010 and 1100 $cm^{-1}$ and it is associated to different overlapped vibrational modes (1027 $cm^{-1}$ wagging of $CH_3$, 1057 $cm^{-1}$ antisymmetric stretching of CCO and 1089 $cm^{-1}$ rocking of $CH_3$ [23]. Finally the isopropanol peak occurs at 953 $cm^{-1}$ and it is assigned to the $CH_3$ rocking mode [24]. These bands were chosen because they are relatively intense and well resolved, and do not show significant overlaps with components of other compounds, a condition that is the prerequisite for the spectroscopic analysis of mixtures.

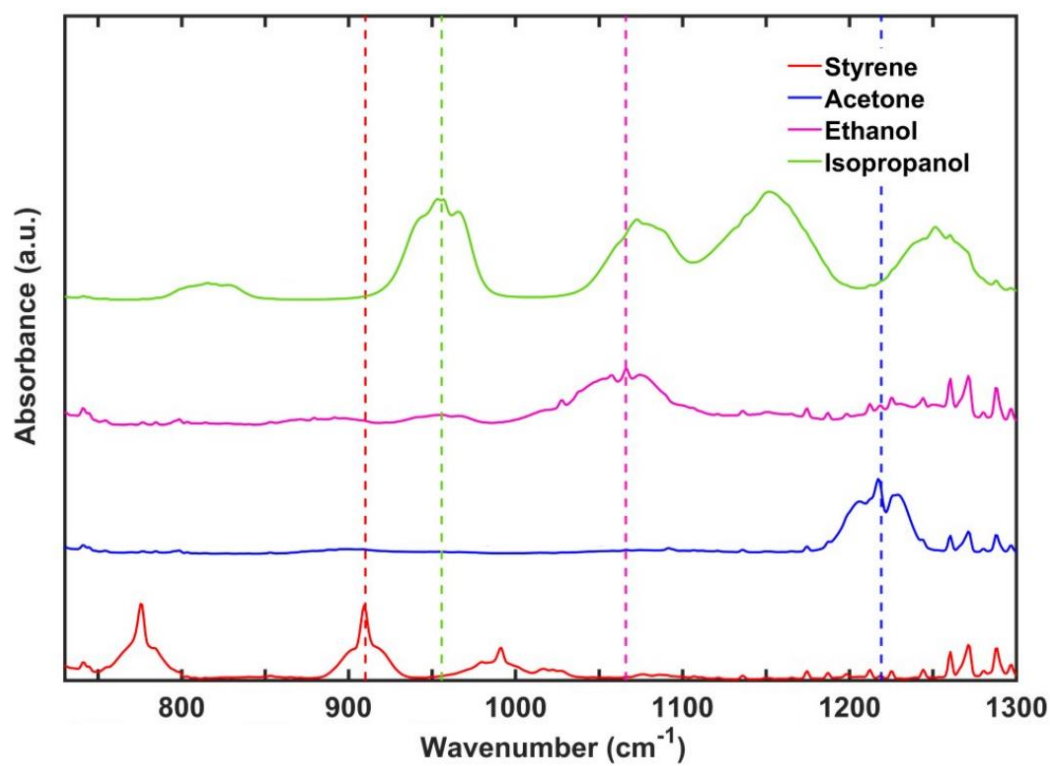

**Figure 2.** FTIR spectra of the studied VOCs in the MIR spectral region ranging between 750 - 1300 cm$^{-1}$. Coloured dashed lines highlight the absorbance peaks considered for the evaluation of the calibration curves of each VOC: styrene (red), acetone (blue), ethanol (pink) and isopropanol (green).

In transmission IR spectroscopy, the absorbance of a specific band is related to the amount of the target molecule via the Beer-Lambert relationship [18]:

$$A(v) = l\, C\, \varepsilon(v) \quad (1)$$

where $A(v)$ is the absorbance (adimensional), $l$ is the optical path of the cell (in cm), and C the concentration of the molecule (in ppm$_v$) and $\varepsilon(v)$ (ppm$_v^{-1}$ cm$^{-1}$) is the absorption coefficient.

Integrated absorbances ($A_i$) were obtained by integrating the area of the characteristic absorption bands for each VOC and averaging over ten spectra recorded sequentially. The uncertainty on the integrated absorbance is the standard deviation calculating on the repeated measurements at the same ppm$_v$ concentration The VOCs concentrations in the multipass gas cell, expressed in ppm$_v$, were calculated as reported in Supporting Information. The uncertainty on the concentration measurements is established to be 1 ppm$_v$ from the PID manufacturer and it is properly scaled following the error propagation equation.

*3.2. Curve calibrations for styrene and individual interfering VOCs*

We obtained the calibration curves for the four VOCs by correlating the integrated absorbance over pathlength for the characteristic VOC peak ($A_i$ was normalized with respect to the multipass optical path of 10 m) with the ppm$_v$ estimated for each gas according to equation (2). The relationship between absorbance vs. ppm$_v$ is expressed by the equation (1) and the resulting curves are displayed in Figure 3, where experimental data are linearly fitted by using the ORIGIN PRO™ software. The extrapolated fitting parameters, including slope and intercept of the calibration lines, are given in Table 2. The slope provides the absorption coefficient of the specific normal mode excited through IR radiation, whereas the intercept, likewise the integrated zero-concentration absorbance, is expected to approach zero.

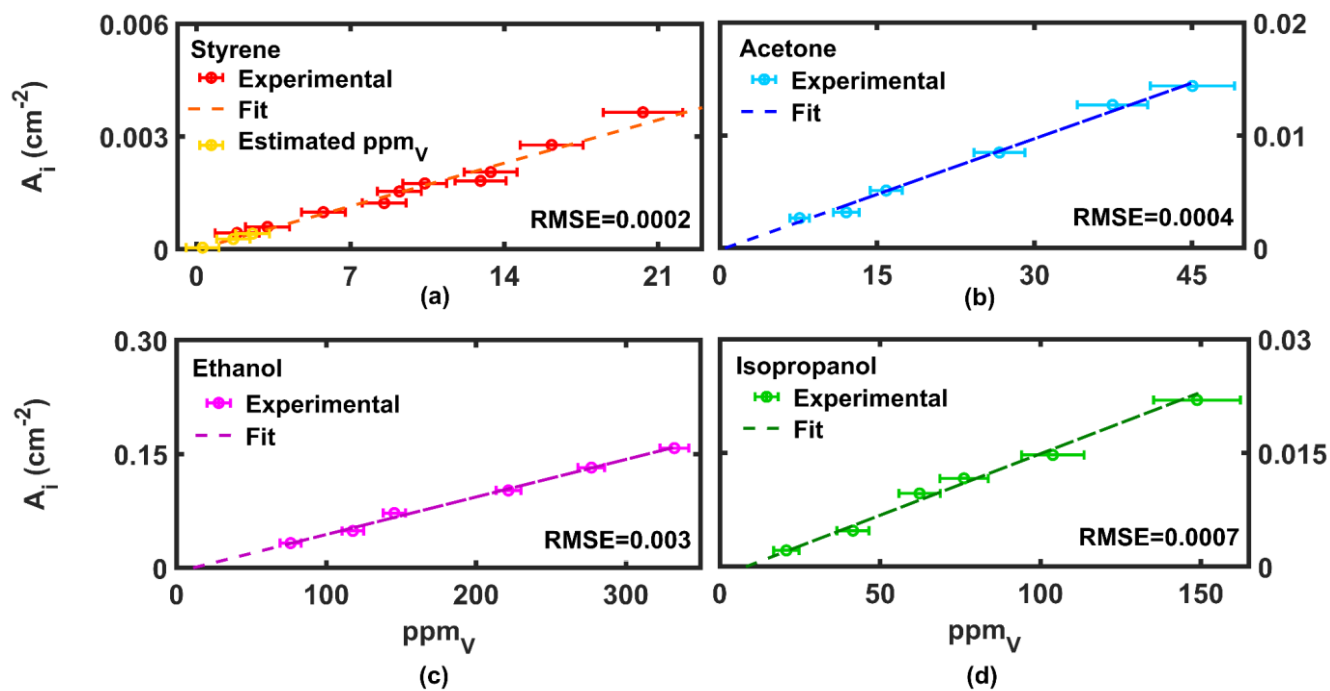

**Figure 3.** Calibration curves for styrene (**a**), acetone (**b**), ethanol (**c**) and isopropanol (**d**), referred to the integrated absorbances $A_i$ vs. ppm$_v$. The experimental data are indicated by points and the fit curves by the dashed lines. In the panel (**a**), we reported the $A_i$ related to sub-ppm$_v$ concentrations in yellow, indirectly estimated. The root-mean-square error (RMSE) is used for the estimation of the differences between experimental values and the adopted linear model.

To calculate the integrated absorbances we selected the bands at 910 cm$^{-1}$ for styrene, at 1229 cm$^{-1}$ for acetone, at 1100 cm$^{-1}$ for ethanol and at 953 cm$^{-1}$ for isopropanol. The goodness of each fit is estimated through the root-mean-square error parameter (RMSE), reported in the corresponding panel of each calibration line. The inspection of Table 2 shows that the most intense absorption is obtained for ethanol, however the absorption coefficients (slopes) associated to the different excited normal modes are all comparable. In addition, we have to underline here that the intercepts are all very close to zero, thus providing a good confirmation to the validity of the calibration.

**Table 2.** Slope and intercept associated with the linear fit of styrene, acetone, ethanol and isopropanol.

| VOCs | Slope (cm$^{-2}$) | Intercept (cm$^{-2}$) |
|---|---|---|
| Styrene | $(1.6 \pm 0.1) \times 10^{-4}$ | $(-0.003 \pm 0.122) \times 10^{-3}$ |
| Acetone | $(3.3 \pm 0.2) \times 10^{-4}$ | $(-0.2 \pm 0.4) \times 10^{-3}$ |
| Ethanol | $(5.0 \pm 0.2) \times 10^{-4}$ | $(-5.5 \pm 3.7) \times 10^{-3}$ |
| Isopropanol | $(1.6 \pm 0.1) \times 10^{-4}$ | $(-1.3 \pm 0.7) \times 10^{-3}$ |

As far as styrene is concerned, we tried to measure ppm$_v$ values near the lower limit of PID detectability. Volumes of the order of μl of liquid styrene were put in the expansion chamber. Exploiting the capability of the multipass cell, we collected IR spectra and computed their corresponding integrated absorbances. Using the calibration line (shown in Figure 3a), we extrapolated the corresponding ppm$_v$ values, which are: 0.28 ± 0.74, 1.67 ± 0.75, 2.55 ± 0.77, well below the PID sensitivity and working linearity.

*3.3. Comparison with laboratory detection setups*

After the calibration of the system, as explained above, we compared the results obtained using the multipass cell with results of experiments performed with both a benchtop and a portable FTIR spectrometer (Figure S1) available from [11][18]. Figure 4 displays the calibration curves of styrene (Figure 4a), acetone (Figure 4b), ethanol (Figure 4c) and isopropanol (Figure 4d) obtained by using all three set-ups: benchtop, portable FTIR and multipass.

For the comparison we display the integrated absorbance, normalized with respect to the different optical path associated with each experimental setup, i.e., 27 cm for the benchtop system, 7 cm for the portable device and 1000 cm for the multipass layout.

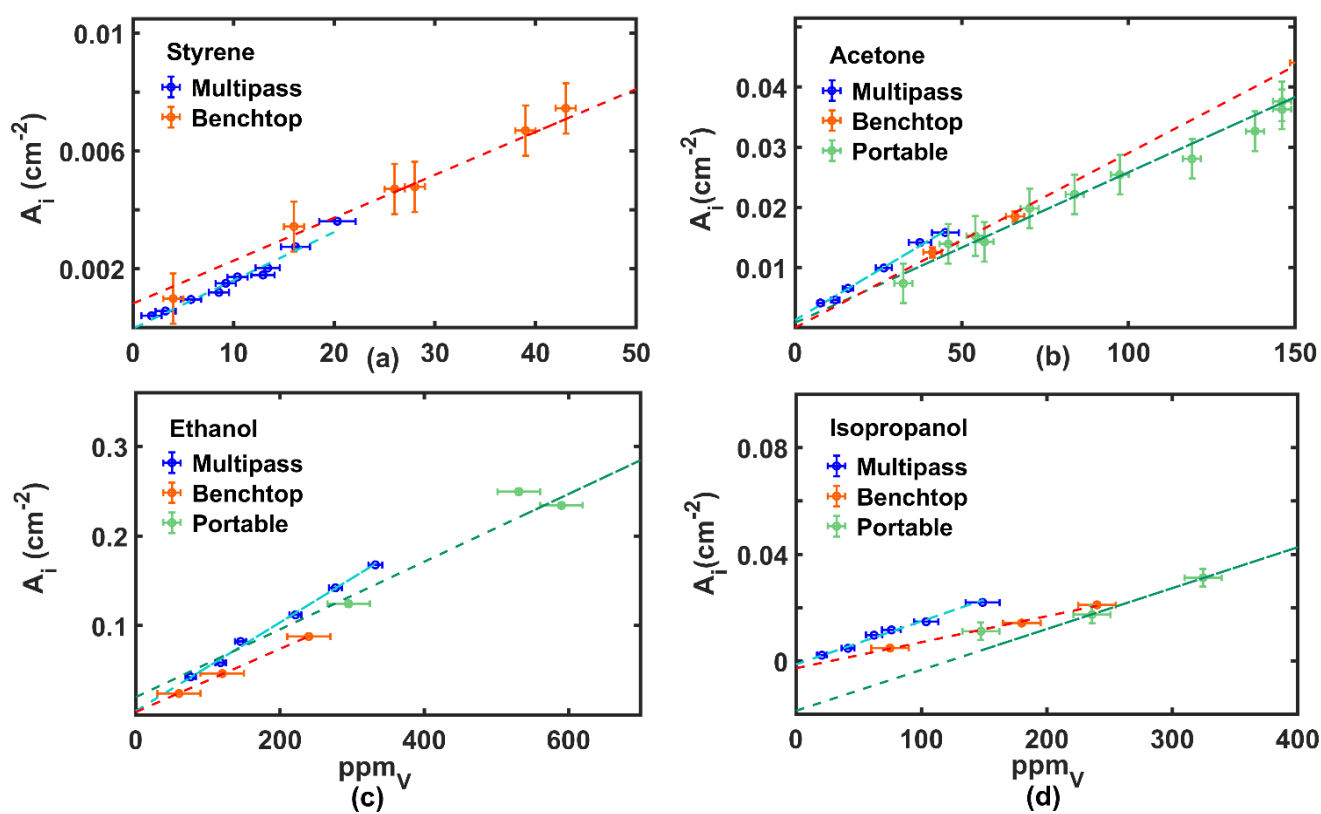

**Figure 4.** Enlargement view of the calibration of the integrated absorbance, normalized at the optical path, vs. ppm$_v$ for styrene (**a**), acetone (**b**), ethanol (**c**) and isopropanol (**d**) obtained with the multipass (light blue line), benchtop (red line) and the portable device (green line), respectively. The whole calibration curves of the integrated absorbance vs. ppm$_v$ obtained with the three setups are reported in figure S3 of the Supporting Information.

For the sake of comparison, either slope and intercept of the calibration curves need to be considered. For any VOC, the slope is expected to be the same, within the uncertainty for any experimental setup, because the absorption coefficient is independent of the specific experimental device or layout employed to perform the experiment.

The plots of Figure 4 show that the intercept relative to the multipass is systematically very close to zero indicating that this device indeed provides data with better sensitivity.

This behavior is easily explained by considering that, even at low VOCs concentration, with the multipass setup, thanks to the longer optical path length absorption bands (see Eq. 1), are more intense and the signal-to-noise ratio (S/N) is higher, ensuring a more accurate measurement.

*3.4. Quantitative analysis of binary and ternary mixtures of VOCs*

One of the main advantages of using IR spectroscopy for VOCs detection is the possibility to discriminate with a high degree of confidence different chemical species, thanks to their different spectral features (see Figure 2), and to extrapolate their relative concentration in ppm$_v$ from their integrated absorbance at the characteristic peaks, using the calibration curves (paragraph 3.2). We tested the detectability of gas compositions, on both binary and ternary mixtures. We prepared three binary solutions mixing volumes of acetone/ethanol (60:40 volume ratio), styrene/acetone (70:30) and styrene/ethanol (70:30), and one ternary solution mixing styrene/acetone/ethanol (45:35:20). We injected the liquid solution into the expansion cell, and waited for the equilibrium condition (liquid-vapor) by monitoring the gas concentration using the PID sensor.

When the interaction among components can be considered negligible, the net absorbance of a gaseous mixture can be considered as the sum of the contribution of the linear absorbance of each component [25-26]. This is known as the multiple absorbers approach

$$A_{mix}^{exp} = \sum_i A_i \qquad (2)$$

where $A_{mix}^{exp}$ is the experimental mixture absorbance and $A_i$ is the estimated absorbance of pure components, due to each species i, at the same pressure and temperature conditions. Observing the Figure 5a, the mixtures of styrene/acetone and acetone/ethanol show absorption bands clearly distinguishable and separated, so we simply integrated the area of their characteristic peaks, as discussed above. On the other hand, styrene and ethanol show non negligible peak overlaps in the 840-920 cm$^{-1}$ and 980-1010 cm$^{-1}$ ranges, so it was not possible to simply integrate in the frequency window of interest for that VOC. It was necessary to fit the mixture spectrum in order to extrapolate the single VOC contribution to the absorbance. We look at the mixture spectrum as the sum of individuals, each one multiplied by a factor representing the VOC concentration

$$X mixture = \alpha \, X ethanol + \beta \, X styrene \qquad (3)$$

Using the Matlab function *fminsearch* we found the most proper α and β coefficients to satisfy this equation (see SI for mathematical specifications). We performed this calculation separately for the two frequency ranges, 830-945 cm$^{-1}$, where styrene referring peak is present, and 946-1161 cm$^{-1}$, where there is the ethanol one, therefore obtaining $\alpha$ from the second one and $\beta$ from the first one.

Mixture spectra are shown in Figure 5(a). Coloured areas are the ones integrated with the single-peak method already discussed in the calibration sections, while the range 830-1160 cm$^{-1}$ for the ternary and the styrene+ethanol mixtures has been analyzed with the fitting procedure and results are reported in Figure 5(b). In Table 3 we report the concentrations in ppm$_v$ for the individual VOCs of the four mixtures, obtained from the integrated areas and the calibration curves.

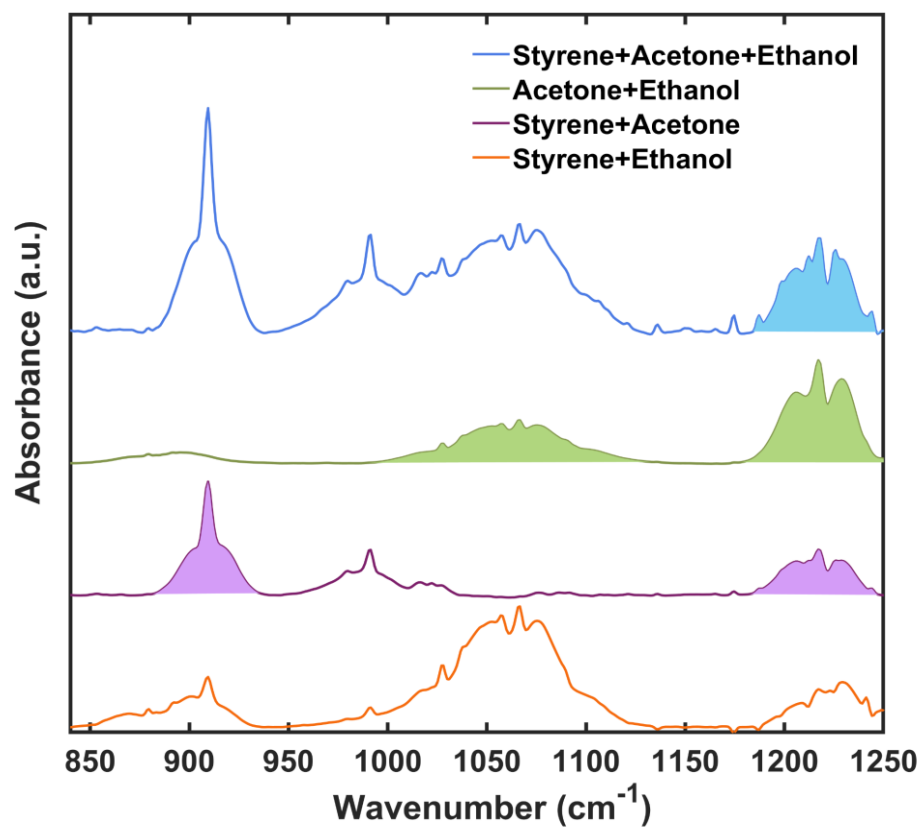

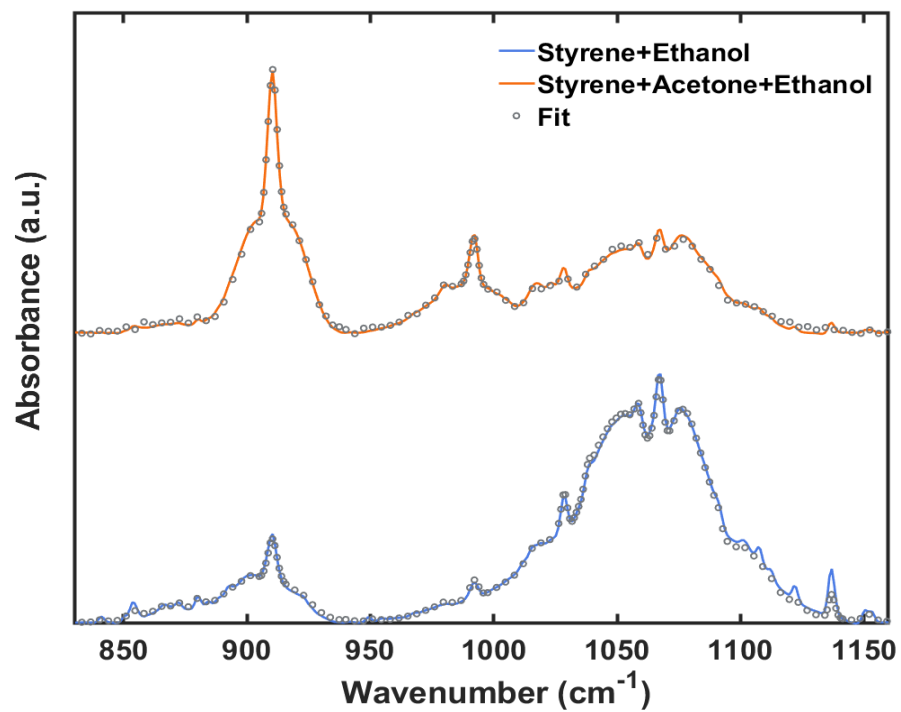

**Figure 5.** (a) Spectra of mixtures. Coloured areas are calculated with the integration method of the OPUS™ 8.2 software. (b) Fit of styrene and ethanol absorbance peaks in the range 830-1160 cm$^{-1}$ for a mixture of styrene+ethanol (orange) and for a ternary mixture (blue).

|  | **Styrene (ppm$_v$)** | **Ethanol (ppm$_v$)** | **Acetone (ppm$_v$)** | **Sum\* (ppm$_v$)** | **PID lecture (ppm$_v$)** |
|---|---|---|---|---|---|
| styrene + ethanol | (3.8 ± 0.8) | (27.2 ± 0.2) | // | (4.7 ± 1.0) | (5.3 ± 0.5) |
| styrene + acetone | (4.90 ± 0.80) | // | (10.3 ± 1.2) | (8.7 ± 1.3) | (11.1 ± 1.0) |
| acetone + ethanol | // | (20.2 ± 7.4) | (27.1 ± 2.1) | (10.5 ± 1.0) | (7.4 ± 0.9) |
| styrene + acetone + ethanol | (6.5 ± 0.9) | (14.3 ± 7.4) | (5.6 ± 1.3) | (9.1 ± 1.6) | (9.9 ± 0.9) |

*\*Sum is intended to be calculated summing up the concentrations scaled using the styrene calibration, so to make possible the comparison with the initial PID lecture, i.e., Sum = ∑ (Concentration (ppm$_v$) / CF) .*

**Table 3.** Concentration values (ppm$_v$) extrapolated from the integrated areas of the absorbance peaks for different mixtures. The sum of the calculated values can be compared with the PID lecture.

The sum is calculated adding the concentrations in ppm$_v$ as extrapolated from the integrated areas, each one divided by the relative CF. PID lecture instead is the concentration in ppm$_v$ measured directly by the instrument. Inspection of the final results (last two columns in Table 3) shows an excellent agreement between the concentrations obtained by summing the contribution of the individual VOCs and the PID readings that represents the total (unresolved) gas concentration. Two values result to be coherent, showing the goodness of fit procedure and of the multipass setup capability to determine mixture components. The data in table 3 thus provide good evidence for the reliability of calibration used to extrapolate the single ppm$_v$ from the IR spectrum collected on the mixture.

**4. Conclusions**

In the present work we provide a new, improved calibration curve for the IR spectroscopic detection for a series of gaseous compounds of environmental and occupational interest. The used technique is based on the combination of an IR Michelson interferometer and a multipass gas cell. In particular, we have calibrated the quantitative analysis in the MIR spectral region for four different VOCs (styrene, acetone, ethanol and isopropanol) focusing on its advantages compared to portable and benchtop devices used previously [11,18], in terms of both detection sensitivity and accuracy. The longer optical path (10 m) inside the multipass cell ensures an excellent S/N ratio, thus allowing the discramination of less intense vibrational bands, overcoming the limitations affecting the VOCs quantification when using conventional set ups.

Styrene, acetone, ethanol and isopropanol were investigated individually in order to obtain accurate calibration curves. The four gas-phase VOCs exhibit good linear relationships between the integrated absorbance of the specific vibrational bands and the corresponding concentrations provided by the independent measurement using a styrene-calibrated PID. The use of the calibration curves provides a reliable method to get sub-ppm$_v$ concentrations well below the PID detection capability, as observed, in particular, for styrene. In order to mimic real "conditions" in indoor environments, binary and ternary VOCs mixtures were also prepared and analyzed. Results proved the capability of this new setup to discriminate the different chemical species present simultaneously in the atmosphere, thanks to their univoque spectral features. In particular, using a mixture, whose total concentration falls within the PID detection range, this spectroscopic approach allows extracting the concentration of individual components otherwise not measurable with the PID sensor. The results of this study paves the way toward the design of improved compact (and portable) air-quality monitoring systems capable of discriminating VOCs present simultaneously in the atmosphere and quantifying them down to the sub-ppm$_v$ range.

**Supplementary Materials:** The following supporting information can be downloaded at: www.mdpi.com/xxx/s1.

**Author Contributions:** "Conceptualization, G.D.V., F.R., S.L., A.D., M.C.P.,G.T. and T.M.; methodology, F.R., M.C.P., T.M., A.D. and S.M.; validation, M.C.P. and T.M.; formal analysis, M.C.P., and T.M.; investigation, F.R., M.C.P., T.M. and A.D.; data curation, M.C.P. and T.M.; writing—original draft preparation, M.C.P., T.M., A.D..; writing—review and editing, G.D.V., S.L., S.M., G.T. , F.R., M.P. and A.M.; visualization, M.C.P., T.M., F.R. and A.D.; supervision, G.D.V. and S.L.; project administration, X.X.; funding acquisition, G.D.V., S.L., G.T., M.P. and A.M. All authors have read and agreed to the published version of the manuscript."

**Funding:** This research was supported by the Ph. D. funding in the framework of "Programma Operativo Nazionale (PON) - Ricerca e Innovazione 2014-2020 - Azione IV.5 - Dottorati su tematiche green. We thank for funding BRIC-INAIL project ID07; NATO Science for Peace and Security Programme under grant No. G5889 – "SARS-CoV-2 Multi-Messenger Monitoring for Occupational Health & Safety-SARS 3M", LazioInnova "Gruppi di Ricerca 2020" of the POR FESR 2014/2020 - A0375-2020-36651 project entitled "DEUPAS -DEterminazione Ultrasensibile di agenti PAtogeni mediante Spettroscopia". Financial support by the Grant to Department of Science, Roma Tre University (MIUR-Italy Dipartimenti di Eccellenza, ARTICOLO 1, COMMI 314-337 LEGGE 232/2016) is gratefully acknowledged.

**Data Availability Statement:** In this section, please provide details regarding where data supporting reported results can be found, including links to publicly archived datasets analyzed or generated during the study. Please refer to suggested Data Availability Statements in section "MDPI Research Data Policies" at https://www.mdpi.com/ethics. If the study did not report any data, you might add "Not applicable" here.

**Conflicts of Interest:** "The authors declare no conflict of interest."

# High Sensitivity real-time VOCs monitoring in air through FTIR Spectroscopy using a Multipass Gas Cell Setup

Annalisa D'Arco[1,2], Tiziana Mancini[2,3], Maria Chiara Paolozzi[4], Salvatore Macis[2,3], Augusto Marcelli[2,5], Massimo Petrarca[3,6], Francesco Radica[7], Giovanna Tranfo[8], Stefano Lupi[2,3] and Giancarlo Della Ventura[4,9,*]

[1] INFN-LNF Laboratori Nazionali Frascati, Via E. Fermi 54, 00044 Frascati, Italy augusto.marcelli@lnf.infn.it, annalisa.darco@roma1.infn.it
[2] Department of Physics, University of Rome 'La Sapienza', P.le A. Moro 2, 00185, Rome, Italy tiziana.mancini@uniroma1.it, stefano.lupi@uniroma1.it, salvatore.macis@uniroma1.it
[3] National Institute for Nuclear Physics section Rome1, P.le A. Moro 2, 00185, Rome, Italy stefano.lupi@roma1.infn.it
[4] Department of Science, University Rome Tre, V.le G. Marconi 446, 00146, Rome, Italy giancarlo.dellaventura@uniroma3.it paolozzi.1698739@studenti.uniroma1.it
[5] Rome International Centre for Materials Science Superstipes, Via dei Sabelli 119A, 00185, Rome, Italy
[6] SBAI, Department of Basic and Applied Sciences for Engineering, University of Rome 'Sapienza', Via Scarpa 16, 00161 Rome, Italy massimo.petrarca@uniroma1.it
[7] Department of Engineering and Geology, University G. d'Annunzio Chieti-Pescara, francesco.radica@unich.it
[8] Department of Occupational and Environmental Medicine, Epidemiology and Hygiene, INAIL, Monte Porzio Catone, 00144 Rome, Italy g.tranfo@inail.it
[9] INGV, Via di Vigna Murata 605, 00143, Rome, Italy

## Supporting Information

S1: PORTABLE AND BENCHTOP SETUPS

S2: PPMV DEFINITION AND DISTRIBUTION

S3: COMPARISON WITH LABORATORY DETECTION SETUPS

S4: FITTING PROCEDURE IN MATLAB

## S1: Portable and benchtop setups

Calibration benchtop Fourier-transform infrared (FTIR) measurements were performed using a Bruker Vertex 70 interferometer. The spectrometer was coupled with a commercial Photo-Ionization Detector (PID) (TA-2100 Styrene Detector from Mil-Ram Technology, Inc., www.mil-ram.com) calibrated for the detection of styrene in the range 10-100 ppm$_v$. According to the manufacturer, this sensor has a sensitivity of 1 ppm$_v$. A schematic drawing of the experimental set-up is shown in Figure S1a. The PID system was mounted on top of the spectrometer chamber and was connected to an external computer for data collection and analysis; the sample compartment is separated by IR transparent windows from the rest of the optical system. Selected amounts of liquid VOCs were placed on a Corning® Petri dish located at the base of the spectrometer sample chamber (IR light path length 28 cm) and allowed to evaporate completely. The gas concentration in the chamber was monitored simultaneously with the PID on top, every 1 s, and by collecting FTIR spectra in continuous mode, every 30 s for the first hour and then every 60 s. Each experiment lasted from 1 to 3 hours. FTIR spectra were obtained averaging 32 scans, with a nominal resolution of 4 cm$^{-1}$ in the 400-5000 cm$^{-1}$ spectral range.

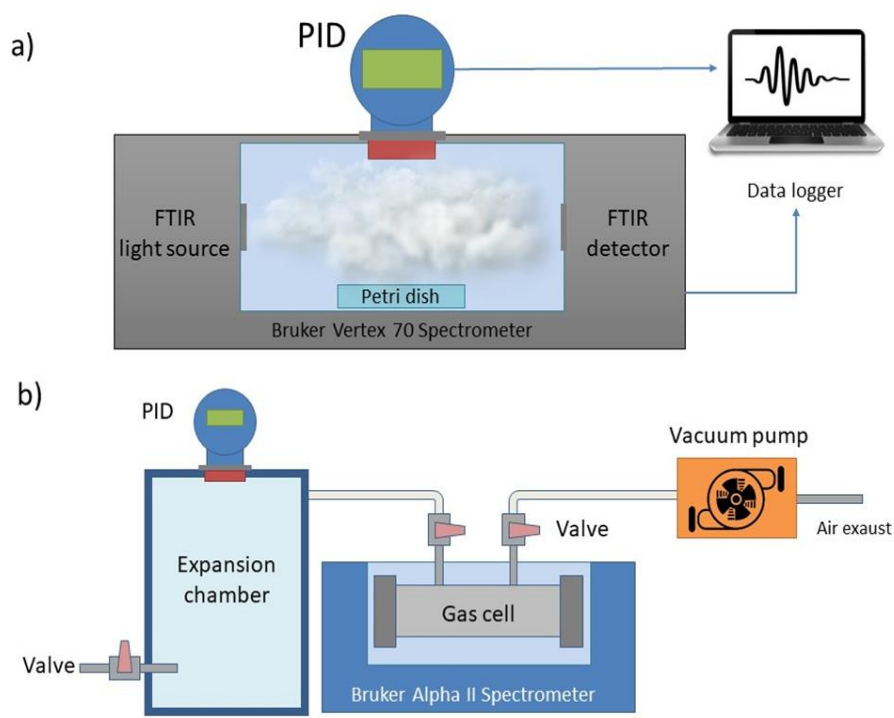

**Figure S1.** Schematic view of the benchtop (**a**) and portable (**b**) setups used for the calibration experiments and reported in our previous work [18].

The collection of each spectrum required around 30 s. Considering the main objective of this work, we selected several common interfering VOCs associated with styrene in workplaces to be tested alone or mixed with styrene. Detection of interfering VOCs was performed using the same PID optimized for styrene. Therefore, the measured values (i.e., equivalent ppm$_v$) were converted into the real concentrations of these gasses using appropriate correction factors (for a PID UV lamp at 10.6 eV). For the calibration of the portable device, we used a Bruker Alpha II spectrometer equipped with a gas cell with a path length of 7 cm (cell volume ~ 0.01 l). The gas cell was connected to a sealed evaporation chamber, where the PID sensor was also installed for real-time monitoring of evaporated VOC. Figure S1b shows a schematic layout of this configuration. For these experiments, different amounts of liquid ethanol, acetone and isopropanol were introduced with a pipette inside the evaporation chamber (chamber volume ~ 0.6 l), and the concentration monitored against time using the PID sensor. As soon as the PID readings indicated the complete evaporation of the liquid within the chamber, the gas was transferred into the gas cell of the spectrometer. The flow was ensured by pre-evacuating the cell (vacuum pressure around 50 mbar) closed by two 5 mm thick KBr windows, using a vacuum pump (Figure S1b). Spectra were collected averaging 16 scans, with a nominal resolution of 4 cm$^{-1}$ in the 4500-500 cm$^{-1}$ range, at the scan velocity of 7.5 kHz. The reduced number of scans with respect to those used for the measurements with the Vertex allowed to keep constant the time required to collect a spectrum (30 s).

In order to compare detection sensitivity of the three experimental setups employed in this work and in the previous one [18], Figure S2 displays three spectra of roughly the same quantity of ethanol gas (around 200-300 ppm$_v$) collected with a conventional benchtop device (in orange), a portable device equipped with a short-path gas cell (in green) and a benchtop device equipped with a multipass cell (in blue). On one hand, the most intense peak around 1050 cm$^{-1}$ is very distinguishable and has the same shape for the three setups. On the other hand, the less intense peak around 875 cm$^{-1}$ is well resolved only in the gas cell, revealing that, as expected, the multipass spectrum is a couple of orders of magnitude more intense than the others.

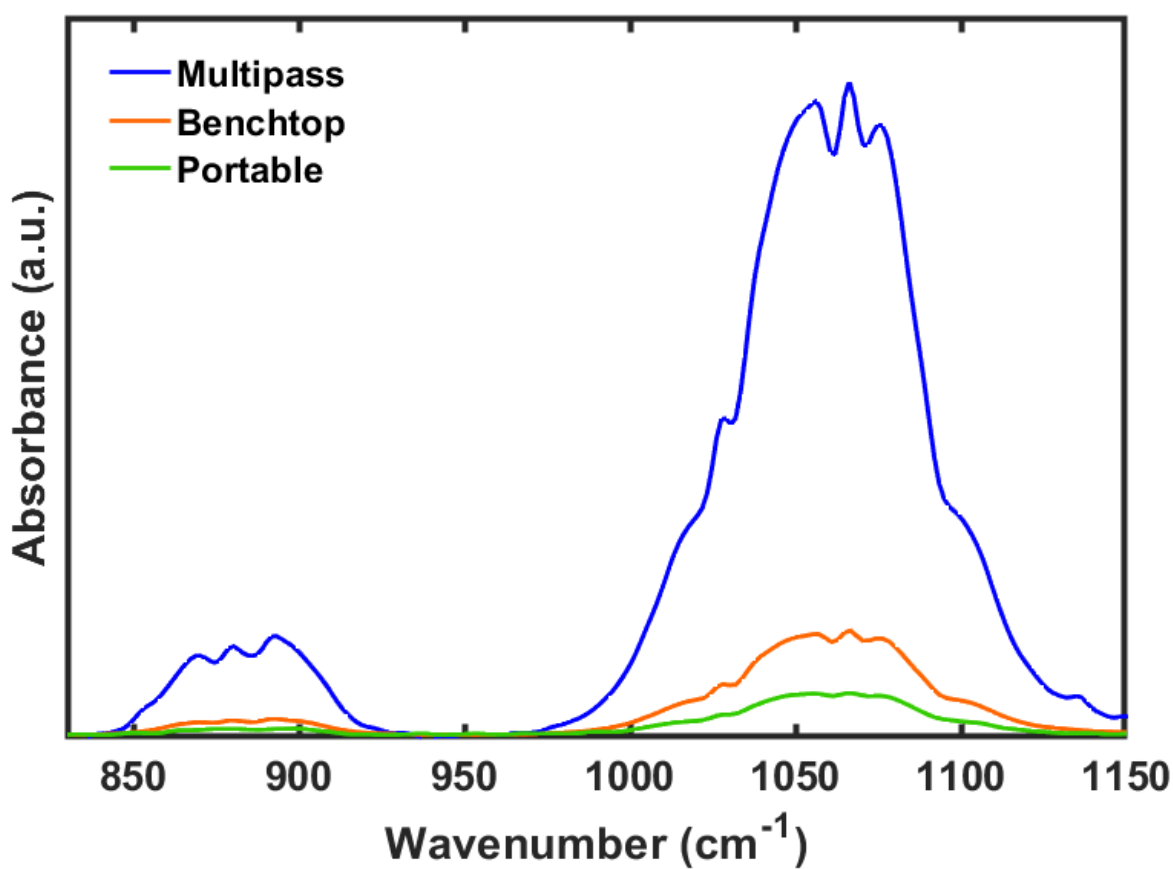

**Figure S2.** Ethanol FTIR spectra in the 800-1150 cm$^{-1}$ spectral range collected using a conventional benchtop device (28 cm path, in blue), a benchtop device equipped with the multipass cell (10 m path, in orange), and a portable device equipped with a short-path gas cell (7 cm FTIR path, in green).

## S2: $ppm_V$ definition and distribution

PID reading is expressed in units of ppm$_v$, i.e. parts per million by volume, related to molar concentration (in mol/m$^3$) as follows:

$$Conc\ (ppm_v) = 24.46 \times Conc\ (mg/m^3) / molecular\ weight\ (g/mol) \quad (1)$$

where 24.46 is air molar volume at 25°.

The PID readings in ppm$_v$ at the beginning of the experiment are referred to the gas concentration in the volume of the evaporation chamber ($V_1$). When the evaporation chamber is connected to the multipass gas cell, the gas expands in the new volume ($V_1+V_2$), as shown in Figure 1. Thus the concentration of the gas molecules is evaluated as follows:

$$C = ppm_{V1} \cdot CF \cdot V1 / (V1 + V2) \quad (2)$$

where ppm$_{v1}$ is the PID reading at the beginning of the experiment, $V_1$ is the volume of the evaporation chamber, $V_2$ is the volume of the multipass gas cell and CF is the factor used to convert the concentration provided by the PID and referred to styrene (the PID we used was factory-calibrated for styrene) to the concentration of other VOCs. CFs are experimentally measured with the procedure described in the PID handbook [19]. Here CFs are provided for different conditions of temperature, humidity and detector voltage. In this work we used the following CFs: 2.75 for acetone, 30 for ethanol, 15 for isopropanol and 1 for styrene (CF from RAE, 2013 for a PID UV lamp at 10.6 eV). The uncertainty on the concentration measurements is established to be 1 ppm$_v$ from the PID manufacturer and it is properly scaled following the error propagation equation.

## S3: Comparison with laboratory detection setups

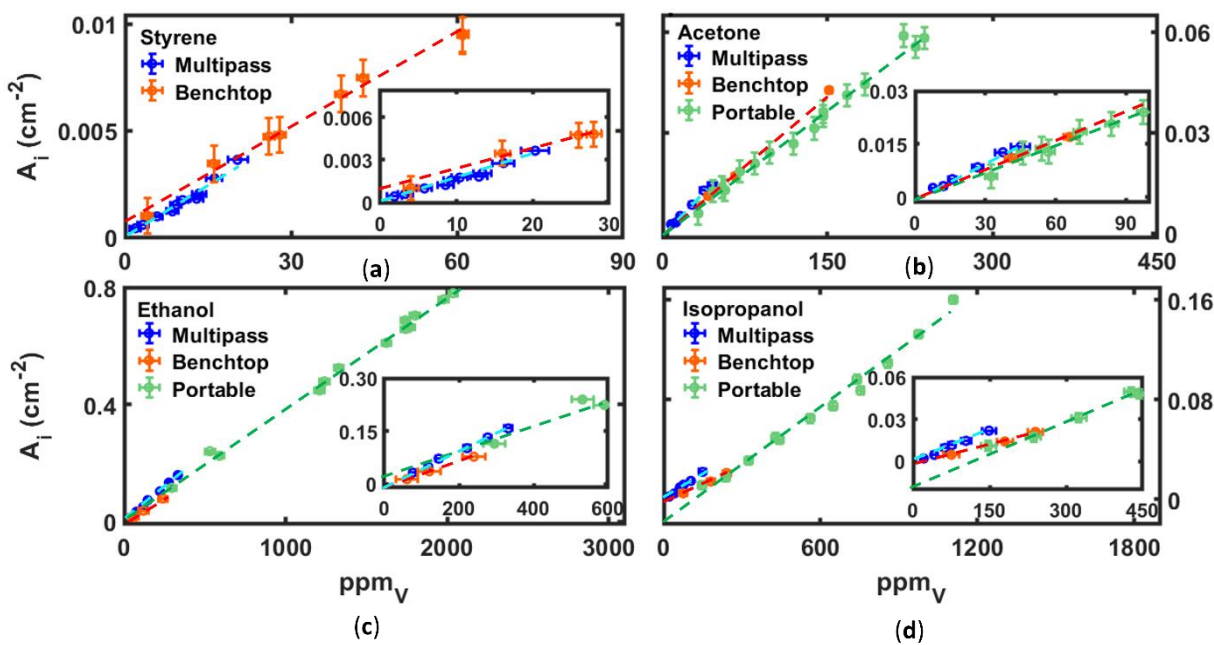

**Figure S3.** Calibration of the integrated absorbance, normalized at the optical path, vs. ppm$_v$ for styrene (**a**), acetone (**b**), ethanol (**c**) and isopropanol (**d**) obtained with the multipass (light blue line), benchtop (red line) and the portable device (green line), respectively. The plots are reported in a larger range compared to the insets, which depict a zoom of each calibration curve, already shown in Figure 4.

## S4: Fitting procedure in Matlab

In VOCs mixtures, in order to discriminate contributions to the whole IR absorbance spectrum of individual elements, an iterative method is employed using the Matlab function *fminsearch*. Looking at the mix spectrum as the linear combination of individual ones, Matlab script find the values of parameters $\alpha$ and $\beta$ which minimize the value of function $f$ defined as follows:

$$f = \sum_i | (X_{mix} - \alpha\, X_{ethanol} - \beta\, X_{styrene}) | \tag{3}$$

where $X_{mix}$, $X_{ethanol}$ and $X_{styrene}$ are the arrays of experimental spectra and the sum is intended on the array index. Procedure is performed separately on two spectral ranges of interest, range 830-945 cm$^{-1}$ and 945-1161 cm$^{-1}$ and therefore styrene and ethanol minimized spectra are obtained.